\documentclass[10pt,aps,prd,twocolumn,showpacs,showkeys,nofootinbib,groupedaddress]{revtex4-1}

\usepackage{graphicx}
\usepackage{dcolumn}
\usepackage{bm}
\usepackage{amsmath}
\usepackage{amsfonts,amssymb}
\usepackage{color}
\usepackage{enumerate}
\usepackage{float}
\usepackage{subfig}
\usepackage{ulem}
\usepackage{times}

\begin{document}

\title{Unveiling cosmography from the dark energy equation of state}

\author{Celia Escamilla-Rivera$^{1}$}
\email{celia.escamilla@nucleares.unam.mx}

\author{Salvatore Capozziello$^{2,3,4}$}
\email{capozzie@na.infn.it}

\affiliation{$^1$ Instituto de Ciencias Nucleares, Universidad Nacional Aut\'onoma de M\'exico, Circuito Exterior C.U., A.P. 70-543, M\'exico D.F. 04510, M\'exico.}
\affiliation{$^2$ Dipartimento di Fisica, University of Napoli ``Federico II'' and Istituto Nazionale di Fisica Nucleare (INFN), Sez. di Napoli, via Cinthia 9, I-80126 Napoli (Italy).}
\affiliation{$^3$ Gran Sasso Science Institute (GSSI), via F. Crispi 7, I-67100, L'Aquila (Italy).}
\affiliation{$^4$ Laboratory for Theoretical Cosmology, Tomsk State University of Control Systems and
Radioelectronics (TUSUR), 634050 Tomsk (Russia).}


\begin{abstract}
Constraining the dark energy equation of state, $w_x(z)$, is one of the main issues of current and future cosmological surveys.
In practice, this requires making assumptions about the evolution of $w_x$ with redshift $z$, which can be manifested in
a choice of a specific parametric form where the number of cosmological parameters play an important role in the observed cosmic acceleration.
Since any attempt to constrain the EoS requires fixing some prior in one form or the other, settling a method to constrain cosmological parameters
is of great importance.  In this paper, we provide a straightforward approach to show how cosmological tests can be improved via a
parametric methodology based on cosmography. Using Supernovae Type IA samplers we show how by performing a statistical analysis of a specific
dark energy parameterisation can give directly the cosmographic parameters values. 
\end{abstract}

\keywords{dark energy; cosmography; modified gravity.}
\pacs{95.36.+x, 04.50.Kd, 98.80.-k}
\maketitle


\section{Introduction}
One of the central challenges of modern cosmology is still to shed light on the physical mechanism behind the cosmic acceleration. 
The interpretation of a handful datasets such observations Supernovas Type Ia (SNeIa), Cosmic Microwave Background Radiation (CMBR), Baryonic Acoustic Oscillations (BAO), among others \cite{Tegmark:2003ud,Jain:2003tba,Riess:2016jrr}, has been useful to constraint the cosmological parameters that define  specific models.  These current observations are consistent with a cosmological constant $\Lambda$, which has its own theoretical and naturalness problems, so it is worthwhile to consider alternatives options to achieve a  cosmic history self-consistent at any epoch. 

A first  scheme is to keep General Relativity (GR) as the standard gravitational theory and modify the stress-energy content of the universe --collectively known as  dark energy and dark matter-- i.e. we modify the right-hand side of the Einstein equations inserting new material ingredients. A second way is to construct a modified gravitational theory, whose additional degrees of freedom can drive the universe acceleration and clustering of structure --called the modify gravity scheme-- via modifying the left-hand side of the Einstein equations, which can lead to new physics on small (ultra-violet) scales and address cosmic dynamics at large (infra-red) scales \cite{review,review2,review3}.

Perhaps the most natural direction to explore the observed cosmic acceleration is at phenomenological level, where both above schemes can be quantified by the evolution of a parameterization (geometric in the case of modified gravity  \cite{Mantica1,Mantica2}) of the dark energy equation of state (EoS)\cite{cardone}. Given the values of the Hubble parameter today, $H_0$, and the current matter density fraction, $\Omega_m$, an arbitrary expansion evolution can be reproduced by assuming a flat Friedmann-Robertson-Walker (FRW) universe with a dark energy component that has an EoS $w_x (z)$. The possible  detection of $w_x(z)$, different from a constant, would be a strong evidence for new gravitational physics and for a new source of cosmic acceleration. Fitting a constant $w_x$ to ongoing data results in good agreement with $-1$, but such fits would have missed subtle variations in the EoS, in specific, if the average happened to be near to $-1$. Alike, using any particular parametric form of $w_x(z)$ is liable for biasing  the outcome. 

 In this line of thought, a kinematical approach can be adequate in order to reveal the correct cosmological paradigm for $w_x(z)$. 
 This \textit{cosmographic} approach
should be able to disclose the fundamental nature
of dark energy without postulating a specific model \textit{a priori} \cite{dunsby1,dunsby2}, but its Achilles' heel lies in the need to be comprehensive by paying particular attention to two issues: the convergence
of the cosmographic series (a problem that can be circumvent by parameterizing cosmological distances with $\zeta$-redshift: $\zeta=z/1+z$) and the relation between the truncation order of the series and the redshift extent of observational data (which can be resolved by analyzing the order of the expansion which maximizes the statistical significance of the fit for a specific dataset\cite{Lazkoz:2013ija}).  As reported in Capozziello  et al. 2019  and references therein\cite{rocco}, these problems could be addressed  considering rational polynomials like Pad\'e and  Chebyshev ones which can improve the convergence of the cosmographic series and  discriminate among viable cosmological models \cite{sen,rocco2}.

The paper is organized as follow: in Sec.\textbf{\ref{sec:background}} we describe the standard background cosmology in order to point out the issues to achieve the path to the cosmography. In Sec \textbf{\ref{sec:constructionEOS}} we show how to construct cosmography from a standard EoS and present the cosmographic parameters in terms of $(H/H_0)^2$. In Sec \textbf{\ref{sec:constructionEOS}} and discuss some bidimensional parameterisations in this landscape. In Sec \textbf{\ref{sec:statistics_analysis}} we perform the statistical analysis for the cosmological models described. Finally, in Sec. \textbf{\ref{sec:remarks}} we will discuss the main results.


\section{Cosmological background}
\label{sec:background}

In general, dynamical dark energy changes the background expansion evolution of the universe with
respect to the standard $\Lambda$CDM model. 
The possible deviation can be tested over
the dark energy EoS parameter $w_x=p_x/\rho_x$, with
$\rho_x$ and $p_x$ the dark energy density and pressure, respectively. Considering the spatial flatness hypothesis we can write the Hubble function as
\begin{equation}\label{eq:friedmann}
E^2(z)=\left(\frac{H(z)}{H_0}\right)^2 =\Omega_{m}(1+z)^3 +(1-\Omega_m)f(z),
\end{equation}
where radiation and curvature contributions have been neglected, and
 $f(z)=\text{exp}\left[3\int^{z}_{0}\frac{1+w_x(\tilde{z})}{1+\tilde{z}}d\tilde{z}\right]$.
In principle, there is not theoretical consensus dictum to select the best $w_x(z)$; however using observational data it is
possible to find parameterisations that are cosmologically viable, e.g. for quiessence models ($w_x=\text{const.}$) we have
$f(z)=(1+z)^{3(1+w_x)}$. If we consider a cosmological constant $w_x=-1$ then $f=1$. 
Forecast for each free parameters $w_i$ present in the models
is  an useful way of comparing the relative performance of different surveys methods in reconstructing the expansion history\cite{Escamilla-Rivera:2016qwv}. 
But, at this point, a generic form of $w_x$ slips out of our hands and we should ask: \textit{Is there a way to write a generic EoS with cosmologically viable characteristics?} And the answer is: \textit{yes.}

The \textit{mare magnum} of gravity theories exhibits background dynamics that can deviate
from our extremely successful descriptive model of the universe --$\Lambda$CDM-- 
which fits a wide number of  observations with great precision, but it is mostly built out from concepts that have so far challenged any proper fundamental physical
understanding --it is enough tricky to try to define the essence of the pillars of this model into the same framework, that is  dark energy, dark matter and inflation--. Moreover, measuring parameters
of the $\Lambda$CDM model with quite precision is not enough if we aspire to an \textit{altissimam scientiam} of our universe covering the whole cosmic history. Therefore, we can:
\begin{itemize} 
\item \textit{develop and test new alternative theories;}
\item \textit{pursue novel observables that can stress $\Lambda$CDM in new, potentially troublesome schemes;}
\item \textit{discover 
anomalies and discordance observations that could expose the weakness of $\Lambda$CDM that might lead us to a self-consistent theory.}
\end{itemize}

Now, if $w_x$ is constant, then
solving several fine-tuning issues and understanding how quantum field theory  vacuum gravitates becomes the main goal to understand the cosmic acceleration. If $w_x$ is \textit{not} constant, this point out to the modifications of GR. 
In a deeper analysis, the EoS also can drop below a value of $-1$, which is said to be in the phantom regime --violating
several energy conditions for a single, minimally coupled scalar field--  hinted to a path  where we need to look for
additional interactions.

Beyond this idea, 
oscillating EoS's  are appealing since they can lighten the coincidence problem, e.g if we match a numerical fit
with the results 
coming from the straightforwardly integration of the field equations derived from the Einstein-Hilbert action with a general function of the Ricci scalar $S=\int f(R)\sqrt{-g}d^4x +S_{\text{matter}}$, it is possible to reproduce \textit{models cosmologically viable}
that can recover $w_x=-1$ at large redshift and allows oscillations in the range of interest for current observations and future surveys. 
At this point \textit{it is already possible to put a theoretical background to parameterisations --and also for alternative gravity models-- at the same level as other parameterisations} \cite{Jaime:2018ftn}.


\section{Constructing cosmography from the equation of state}
\label{sec:constructionEOS}

Motivated by the latter result, a cosmographic study can be performed in order to infer how much dark energy or alternative components are required in regards to satisfy the cosmological  equations. The underlying philosophy of cosmography is to evoke the cosmological principle only, i.e the FRW metric is the only key ingredient that this approach uses for obtaining bounds on the observable universe. However, this approach requires a fiducial model, e.g. we can assume a flat quiessence model or a dynamical dark energy model. To this aim, we can write
an expression
 where we do not impose any form of dark energy by formally solve Eq.(\ref{eq:friedmann}) to obtain
\begin{equation}\label{eq:newcosmoEoS}
1+w(z)= \frac{1}{3}\frac{\left[E(z)^2 -\Omega_m(1+z)^3\right]' (1+z)}{E(z)^2 -\Omega_m (1+z)^3},
\end{equation}
where the prime denotes $d/dz$ and, with the definition of 
$H(z)=\dot{a}/a=H_0 E(z)$, where $H_0$ is the Hubble parameter observed,
we can determine the cosmographic parameters by avoiding to  integrate it directly to get $a(t)$ and consider that 
\begin{equation}
\dot{}=d/dt =-(1+z)H(z)d/dz. \label{eq:physical_time}
\end{equation} 
Using the definitions of the deceleration and jerk parameters, respectively:
\begin{eqnarray}
q(z) \equiv -\frac{a\ddot{a}}{\dot{a}^2} = -1 +\frac{1}{2}(1+z)\frac{[E(z)^2]'}{E(z)^2}, \label{eq:q} 
\end{eqnarray}
\begin{eqnarray}
j(z) \equiv \frac{\dddot{a}a^2}{\dot{a}^3}=\frac{1}{2}(1+z)^2 \frac{[E(z)^2]''}{E(z)^2} -(1+z)\frac{[E(z)^2]'}{E(z)^2} +1, \quad\quad \label{eq:j}
\end{eqnarray}
which are functions that are now solely in terms of the dimensionless Hubble function.
By solving and evaluate them at $z=0$ we get the usual cosmographic series \cite{Capozziello:2011tj}
\begin{eqnarray}\label{eq:Hcosmo}
H(z)&=& H_0 +\frac{dH}{dz}\bigg\rvert_{z=0} z +\frac{1}{2!}\frac{d^2 H}{dz^2}\bigg\rvert_{z=0} z^2 +\frac{1}{3!} \frac{d^3 H}{dz^3}\bigg\rvert_{z=0} z^3 +\ldots, \nonumber\\
\end{eqnarray}
where the extra cosmographic parameters -- the snap $s$ and lerk $l$-- have longer expressions and involve fourth and fifth order derivatives of $H(z)$ in a sequence that can
be expressed as: $\dot{a}=aH, \quad \ddot{a}=-qaH^2, \quad \dddot{a}=jaH^3, \quad \ddddot{a}=saH^4, \quad \text{and} \quad d^5 a/dt^5=laH^5$. Usually, further higher cosmographic parameters in terms of the latter can be expressed as \cite{Luongo:2011zz}:
\begin{eqnarray}
s&=& \frac{H^{(3)}}{H^4} + 4j +3q (q+4) +6, \\
l &=& \frac{H^{(4)}}{H^5} -24-60q -30q^2 -10j (q+2) +5s,
\end{eqnarray}
where the numbers inside parenthesis indicate the third and fourth derivatives with respect to the cosmic time. In order to get the corresponding expressions as in
Eqs.(\ref{eq:q})-(\ref{eq:j}), we can use again Eq.(\ref{eq:physical_time}) and re-express the Hubble parameter as: $dz/dt = (1+z)H_{0} -H(z)$. Similar expressions can be achieved expressing everything in terms of the function $E(z)$ and its derivatives. However the information one can acquire is exactly equivalent. 


\subsection{Recovering $\Lambda$CDM from the cosmographic equation of state}

So far, with the above ideas, the evolution of the Hubble parameters in the cosmography at small redshift $z$ can be expressed as Eq. (\ref{eq:Hcosmo}). However, this kind of standard cosmography has two problems: \textit{(1)} the cosmographic series encounters convergence problems at high redshift and \textit{(2)} with more accuracy in the series, the necessity of more cosmographic parameters increases. The first problem can be solved by performing a simple change of variable over the redshift: $\zeta=z/1+z$. The latter relies in how many terms we need to include to obtain a good precision over the cosmographic parameters, which after constrain them are needed to compute the standard cosmological parameters. 

Our goal is to show how Eq.(\ref{eq:newcosmoEoS}) can help to relax problem \textit{(2)}, in order to found the corresponding constraints over $q$ and $j$ directly by using the best fits obtained for the cosmological parameters with observations. For this, we emphasize that this methodology offers a path from a cosmographic approach to cosmology without consider high order polynomials and dealing with the problem \textit{(1)}. Also, a directly form of $w(z)$ can be achieved once the cosmological parameters are constrained by observations.

\begin{figure}
\centering
\includegraphics[width=0.45\textwidth,origin=c,angle=0]{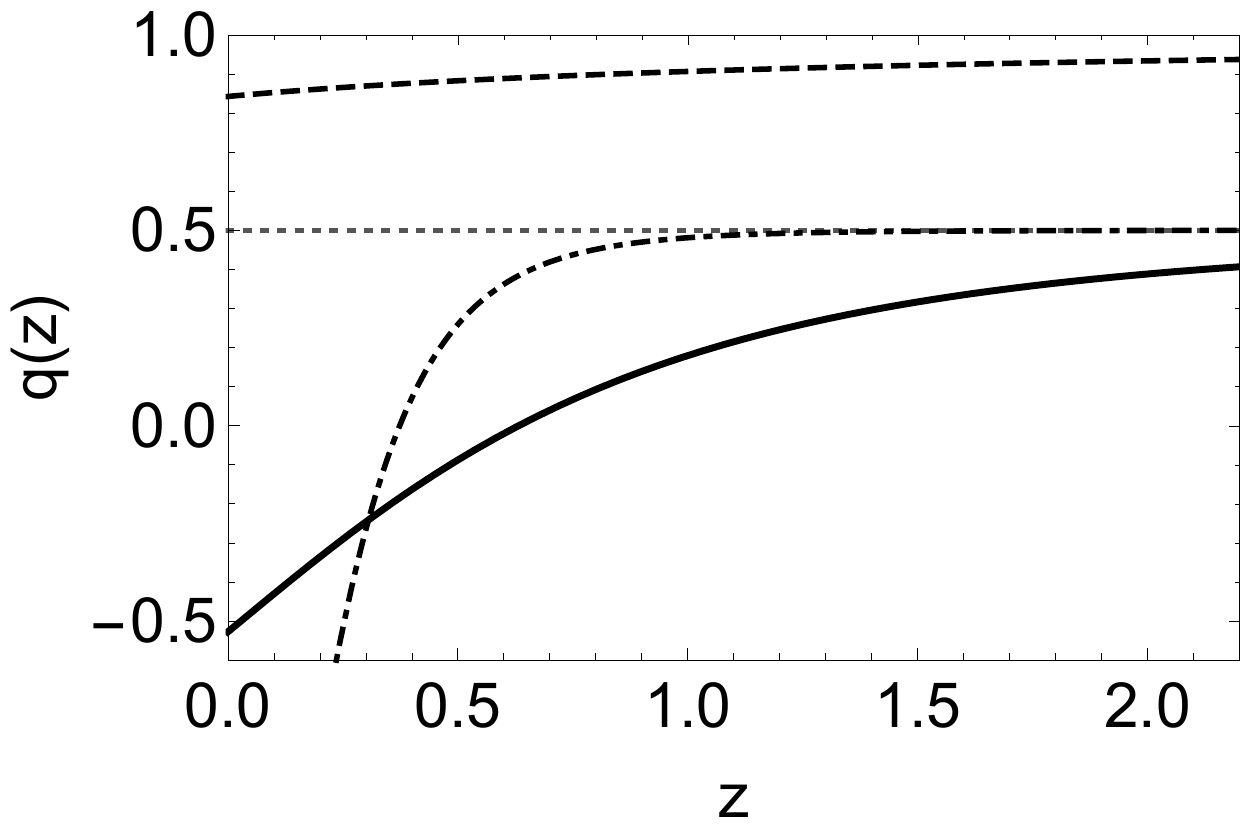}
\includegraphics[width=0.42\textwidth,origin=c,angle=0]{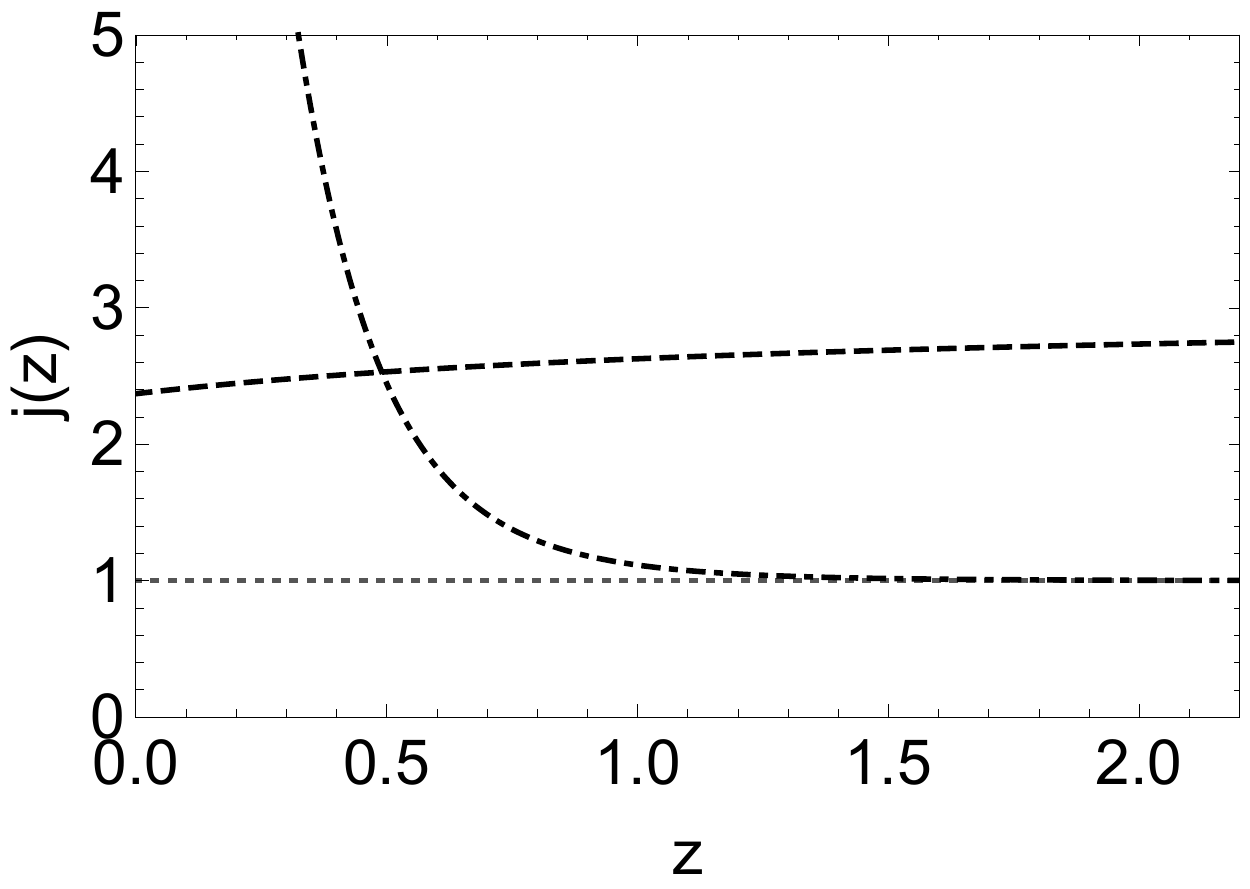} 
\caption{Evolution of Eqs.(\ref{eq:qw})-(\ref{eq:jw}) for different $w$-constant models in the redshift range of interest according to current observations. For instance, we consider a fixed prior $\Omega_m=0.315\pm 0.007$ from Planck 2018 \cite{Aghanim:2018eyx}. The scenarios are: dust model (dotted gray line); $\Lambda$CDM model (solid line) --which for $j$ correspond to the same value for the dust case--; ultra-relativistic model $w=1/3$ (dashed line) and phantom model $w<-1$ (dot-dashed line).} 
\label{fig:evolution_qj_wmodels}
\end{figure}

Let us start with a standard example: a $w$-constant flat cosmological model ($w$CDM) with $E(z)^{2}=\Omega_{m}(1+z)^3 +\Omega_{x}(1+z)^{3(1+w)}$, where $\Omega_{m}$ is the present matter density and $\Omega_{x}=(1-\Omega_{m})$ the dark energy density. According to Eqs.(\ref{eq:q})-(\ref{eq:j}), we can obtain for this model the following cosmographic parameters:
\begin{eqnarray}
q(z)=\frac{\Omega_{m}(1+z)^3 +(1+3w)\Omega_{x}(1+z)^{3(1+w)}}{2\left[\Omega_{m}(1+z)^3 +\Omega_{x}(1+z)^{3(1+w)}\right]} , \label{eq:qw}
\end{eqnarray}
\begin{eqnarray}
j(z)= 1+ \frac{9w(1+w)\Omega_{x}(1+z)^{3(1+w)}}{2\left[\Omega_{m}(1+z)^3 +\Omega_{x}(1+z)^{3(1+w)}\right]}, \label{eq:jw}
\end{eqnarray}
from where we can notice that for dust $w=0$: $q=1/2$ and $j=1$ holds for any redshift. If we consider the evaluation of these parameters at $z=0$ and for a particular value of $w_{x}=-1$ we recover directly the standard $\Lambda$CDM scenario: $3/2(q_0 +j_0)= \Omega_{m}$. Some constant scenarios can be
computed using the above equations as it is show in Figure \ref{fig:evolution_qj_wmodels}.

\subsection {Recovering the cosmographic equation  of state  from $w(z)$ cosmological models}
\label{sec:models}

By using Eqs.(\ref{eq:newcosmoEoS})-(\ref{eq:q})-(\ref{eq:j}), we can easily derive the cosmographic parameters using a specific form of $w(z)$ without dealing with
high derivatives for the solution of the scale factor. Consider the following examples:
\begin{itemize}
\item Linear-Redshift case. The dark energy EoS  for this is given by \cite{Huterer:2000mj,Weller:2001gf}:
\begin{equation}\label{linear}
\begin{aligned}
w(z) = w_0 - w_a z,
\end{aligned}
\end{equation}
which can be reduced to $\Lambda CDM$ model $(w(z)=w=-1)$ for $w_0 =-1$ and $w_a =0$. Inserting Eq. (\ref{linear}) into $f(z)$ from Eq.(\ref{eq:friedmann}), we obtain
\begin{eqnarray}\label{linear p}
 E(z)^2 &=& \Omega_m (1+z)^3
 +\Omega_{x}(1+z)^{3(1+w_0 +w_a)} e^{-3w_a z}. \quad\quad
\end{eqnarray}
However, this ansatz diverges at high redshift and consequently yields strong constraints on $w_a$ in
studies involving data at high redshifts, e.g., when we use CMB data \cite{Wang:2007dg}.
We can compute the cosmographic parameter for this case as: 
\begin{widetext}
\begin{eqnarray}
q=\frac{\Omega _{x} \left(-3 z w_a+3 w_0+1\right) (z+1)^{3 \left(w_a+w_0\right)}+\Omega _m e^{3 
   w_a z}}{2 \left(\Omega _{x} (z+1)^{3 \left(w_a+w_0\right)}+\Omega _m e^{3  w_a z}\right)}, 
   \end{eqnarray}
   \begin{eqnarray}
j&=& \{ \Omega _{x} \left[9 z^2 w_a^2-3 w_a \left(6 w_0 z+4 z+1\right)+9 w_0^2+9 w_0+2\right]
   (z+1)^{3 \left(w_a+w_0\right)}  +2 \Omega _m e^{3 z w_a}\}  
\nonumber\\   && 
\times \left\{2 \left[\Omega _{x} (z+1)^{3\left(w_a+w_0\right)}+\Omega _m e^{3 z w_a}\right]\right\}^{-1} ,
\end{eqnarray}
\end{widetext}
which at $z=0$ correspond to
\begin{eqnarray} 
q_0&=&\frac{\left(3 w_0+1\right) \Omega _{x}+\Omega _m}{2 \left(\Omega _{x}+\Omega _m\right)}, \\
j_0&=&\frac{\left(-3 w_a+9 w_0^2+9 w_0+2\right) \Omega _{x}+2 \Omega _m}{2 \left(\Omega
   _{x}+\Omega _m\right)}.
\end{eqnarray}
Notice that we do not have any dependence of $w_a$ in the cosmographic parameter $q_0$. This rule is preserved by any bidimensional parameterisation.
\item Chevallier-Polarski-Linder (CPL) case. A simple parameterisation that shows interesting properties \cite{Chevallier:2000qy,Linder:2007wa} and, in particular, can be
represented by two parameters that exhibit the present value of the EoS  $w_0$ and its overall time evolution is:
\begin{eqnarray}
 w (z)= w_0 +\left(\frac{z}{1+z}\right) w_a.  \label{eq:cpl}
\end{eqnarray}
The evolution for this parameterisation is given by:
\begin{eqnarray}\label{CPL}
 E(z)^2&=& \Omega_m (1+z)^{3}+\Omega_{x}(1+z)^{3(1+w_0 +w_a)} \nonumber \\ && \times e^{-\left(\frac{3w_a z}{1+z}\right)}.
\end{eqnarray}
Following the above prescription, we can obtain the following cosmographic parameters:
\begin{widetext}
\begin{eqnarray}
q &=&\frac{\Omega _{x} \left(3 z w_a+3 w_0 (z+1)+z+1\right) (z+1)^{3 \left(w_a+w_0\right)}+(z+1) \Omega
   _m e^{\frac{3 z w_a}{z+1}}}{2 (z+1) \left(\Omega _{x} (z+1)^{3 \left(w_a+w_0\right)}+\Omega _m
   e^{\frac{3 z w_a}{z+1}}\right)}, \\
   j&=& [\Omega _{x} \left(9 z^2 w_a^2+3 (z+1) w_a \left(6 w_0 z+3 z+1\right)+\left(9 w_0
   \left(w_0+1\right)+2\right) (z+1)^2\right) 
   \times (z+1)^{3 \left(w_a+w_0\right)}+2 (z+1)^2 \Omega _m e^{\frac{3
   z w_a}{z+1}}] \nonumber \\ && \times
   \left\{2 (z+1)^2 \left(\Omega _{x} (z+1)^{3 \left(w_a+w_0\right)}+\Omega _m e^{\frac{3 z
   w_a}{z+1}}\right)\right\}^{-1}.
\end{eqnarray}
\end{widetext}

\item Redshift squared (R-S) case. This model \cite{Barboza:2008rh} brings a step forward in redshift regions where the CPL parameterisation
cannot be extended to the entire history of the universe. Its functional form is given by:
\begin{eqnarray}
w(z)=w_0 + \frac{z(1+z)}{1+z^2} w_a,
\end{eqnarray}
which is well-behaved at $z\rightarrow -1$. The evolution of this model can be written as:
\begin{eqnarray}\label{BA}
E(z)^2&=&\Omega_m (1+z)^{3}+(1-\Omega_m) (1+z)^{3(1+w_0)} \nonumber \\ && \times (1+z^2)^{\frac{3w_a}{2}},
 \end{eqnarray}
For this case we obtain the following cosmographic parameters:
\begin{widetext}
\begin{eqnarray}
q &=& [\Omega _{x} (z+1)^{3 w_0} \left(z^2+1\right)^{\frac{3 w_a}{2}} \left(3 z (z+1) w_a+\left(3
   w_0+1\right) \left(z^2+1\right)\right) 
   +\left(z^2+1\right) \Omega _m ] 
\nonumber\\ &&  \times \left[2 \left(z^2+1\right) \left(\Omega_{x} (z+1)^{3 w_0} \left(z^2+1\right)^{\frac{3 w_a}{2}}+\Omega _m\right)\right]^{-1}, \quad \quad \\
j&=& \frac{1}{2}(3 \Omega _{x} (z+1)^{3 w_0} [(z+1) w_a \left(6 w_0
   \left(z^3+z\right)+z (z (3 z-1)+5)+1\right)
   +3 z^2 (z+1)^2 w_a^2+3 w_0 \left(w_0+1\right)
   \left(z^2+1\right)^2] +2) \nonumber \\&&
   \times \left\{\left(z^2+1\right)^2 \left(\Omega _m \left(z^2+1\right)^{-\frac{3
   w_a}{2}}+\Omega _{x} (z+1)^{3 w_0}\right)\right\}^{-1}.
\end{eqnarray}
\end{widetext}

\end{itemize}


\section{The analysis by the  observational data}
\label{sec:statistics_analysis}

Cosmographic parameters are important since their character positive (negative) sign immediately show an decelerating (accelerating) cosmic expansion. However, all of them are not observables quantities, therefore we require to perform a data fit over a specific model rewritten in terms of these parameters using astrophysical observations. However, this process results an extension of the two basic problems mentioned: we need to suggest a new variable over the redshift and then found the best fit of the series depending the cosmological priors, then use these values and perform again a fit to found the original cosmological free parameters of the model. 
With our proposal, we can obtain \textit{directly} the cosmographic parameters values by only fitting a specific model without dealing with the mentioned problems, i.e.
we are constraining directly the cosmological parameters for the model and use them to compute the cosmographic parameters $q(z)$ and $j(z)$ (also $q_0$ and $j_0$) without considering high order polynomials \cite{Capozziello:2017nbu} or change of variables over the redshift.
Therefore, through the best fits for each free cosmological parameters in the EoS we can compute Eqs.(\ref{eq:qw})-(\ref{eq:jw}) for a specific dark energy model.

In this paper we are going to consider two different SNeIa samples current available. 
The reason is that the usual methodology to work with the fitting of the cosmographic parameters are based in writing parameterisations of the cosmological distances in terms of the EoS, which carries out Taylor expansion of the redshift drift \cite{Capozziello:2011tj,Zhang:2016urt}. With our proposal we perform a simple fit for a specific EoS and use the results to compute directly $q_0$ and $j_0$ without dealing with a system of several free parameters, which usually consist of: free cosmological parameters (depending the dimension of the model), the traditional $\Omega_m$ and $H_0$ (which can be consider as flat priors if we want a system with less dimensions) and several free (depending of the series) cosmographic parameters. 

In order to develop our proposal, we carry out the following steps:
\begin{enumerate}[(a)]
\item With an astrophysical catalog (see Secs. \ref{ssec:sneia}-\ref{ssec:pantheon}), we perform a numerical fitting of the EoS cosmological parameters.
\item Through a statistical inference analysis, we can choose the best fits values in order to reduce tension over $H_0$.
\item Using the values obtained in step (b), we can compute Eqs.(\ref{eq:q})-(\ref{eq:j}) with the certainty given by the prior distribution of the free parameters for each model and
without extend the priors to any range of redshift beyond the convergence radius of the series. This step is crucial since all the methodologies in the literature impose a minimum requirement to obtain a positive luminosity distance or positive $H^2$ in a redshift range $0<\zeta<1$.
\end{enumerate}
 
\subsection{JLA Type Ia supernovae compilation} \label{ssec:sneia}
To perform the cosmological test we will employ the SNe Ia catalog available: \mbox{the JLA
\cite{Betoule:2014frx}}.
Its binned compilation shows the same trend as using the full catalog itself, for this reason we will use this reduced sample
which can be found in the above reference.
This dataset consist of $N_{\text{JLA}}=31$ events distributed over the redshift interval $0.01< z <1.3$. We remark that the
covariance matrix of the distance modulus $\mu$ used in the binned sample already estimated accounting
various statistical and systematic uncertainties\cite{Betoule:2014frx}.

The distance modules of the \mbox{JLA sample} is given by:
\begin{eqnarray}
\mu(z_i, \mu_0) = 5 \log_{10}\left[(1+z)\int_{0}^{z}{d\tilde{z}E^{-1}(\tilde{z},\Omega_m;w_0,w_a)}\right] +\mu_0,  \nonumber \\
\end{eqnarray}
where $(w_0, w_a)$ are the free parameters of the model. We
can compute the best fits
by minimizing the~quantity
\begin{eqnarray}
\chi_{\text{SN}_{\text{JLA}}}^2
=\sum^{N_{\text{JLA}}}_{i=1}{\frac{\left[\mu(z_i ,\Omega_m ;\mu_0,
w_0,w_a)-\mu_{\text{obs}}(z_i)\right]^2}{\sigma^{2}_{\mu,i}}},
\end{eqnarray}
where the $\sigma^{2}_{\mu,i}$ are the measurements variances.

\subsection{Pantheon Type Ia supernovae compilation} \label{ssec:pantheon}
This sample consist in 40 bins \cite{Scolnic:2017caz} compressed. Notice that, since we are performing EoS's that at some point recover $\Lambda$CDM, the binned catalog is not a problem in the sense of favoring this model.
Type Ia supernovae can give determinations of the distance modulus $\mu$, whose theoretical prediction is related to the luminosity distance $d_L$ according to:
\begin{equation}\label{eq:lum}
\mu(z)= 5\log{\left[\frac{d_L (z)}{1 \text{Mpc}}\right]} +25,
\end{equation}
where the luminosity distance is given in Mpc. In the standard statistical analysis, one adds to the distance modulus the nuisance parameter $M$, an unknown offset sum of the supernovae absolute magnitude (and other possible systematics), which is degenerate with $H_0$. As we are assuming spatial flatness, the luminosity distance is related to the comoving distance $D$ via
\begin{equation}
d_{L} (z) =\frac{c}{H_0} (1+z)D(z),
\end{equation}
where $c$ is the speed of light, so that, using (\ref{eq:lum}) we can obtain
\begin{equation}
D(z) =\frac{H_0}{c}(1+z)^{-1}10^{\frac{\mu(z)}{5}-5}.
\end{equation}
Therefore, the normalised Hubble function $H(z)/H_0$ can be obtained by taking the inverse of the derivative of $D(z)$ with respect to the redshift $D(z)=\int^{z}_{0} H_0 d\tilde{z}/H(\tilde{z})$.
Since we are taking nuisance parameter $M$ in the sample, we choose the respective values of $M$ from a statistical analysis of the $\Lambda$CDM model with Pantheon sample obtained by fixing $H_0$ to the Planck value \cite{Aghanim:2018eyx}. To perform this we have used a MontePython code and obtained a value of $M=-19.63$.

\begin{widetext}
\begin{center}
\begin{table}[]
\caption{Dark energy parameterisations with best fits and cosmographic parameters values at 3-$\sigma$ using JLA SNeIa binned sample.}
{
\begin{tabular}{|c|c|c|c|c|}
\hline
\hline {\bf Model}            	& {\bf Bestfit parameters } 						  & $\bf{q_0}$    		&$\bf{j_0}$  	& Evidence against $\Lambda$CDM \\
\hline $\Lambda$CDM 	&$\Omega_{m}=0.315\pm 0.007$       			  & $ -0.523 \pm 0.011$    &$-1.$ 	  	&	-						\\
\hline Linear 		      	&$w_0= -0.991\pm 0.036, \; w_a= 0.297\pm 0.779 $   & $ -0.333 \pm 0.040 $   &$0.667\mp 0.906$ 	&	1.904 (positive)			\\
\hline CPL 		     	&$w_0= -0.997\pm 0.049, \; w_a= -0.337\pm 1.822 $  & $ -0.524 \pm 0.061 $   &$0.645 \pm 1.726$ & 	1.912 (positive)		\\
\hline R-S 		              	&$w_0= -0.993\pm 0.034, \; w_a= -0.245\pm 0.545 $  & $-0.520 \pm 0.045 $    &$0.727 \pm 0.459$ &	1.921 (positive)		\\
\hline
\hline
\end{tabular}\label{table_bestfits}
}
\end{table}
\end{center}

\begin{center}
\begin{table}[]
\caption{Dark energy parameterisations with best fits and cosmographic parameters values at 3-$\sigma$ using Pantheon SNeIa binned sample.}
{
\begin{tabular}{|c|c|c|c|c|}
\hline
\hline {\bf Model}            	& {\bf Bestfit parameters } 						  & $\bf{q_0}$    		&$\bf{j_0}$  & Evidence against $\Lambda$CDM \\
\hline $\Lambda$CDM 	&$\Omega_{m}=0.321\pm 0.006$       			  & $ -0.566 \pm 0.011$    &$-1.$ 	  &			-			 \\
\hline Linear 		      	&$w_0= -1.091\pm 0.024, \; w_a= 0.311\pm 0.071 $   & $ -0.445 \pm 0.042 $   &$0.986\mp 0.950$ 	&	1.856 (positive)	\\
\hline CPL 		     	&$w_0= -1.064\pm 0.002, \; w_a= -0.179\pm 0.003 $  & $ -0.593 \pm 0.013 $   &$1.025 \mp 0.004$  &	2.153.  (positive)	\\
\hline R-S 		              	&$w_0= -0.995\pm 0.031, \; w_a= -0.145\pm 0.015 $  & $-0.564 \pm 0.040 $    &$0.829 \mp 0.024$ &	3.121 (strong)				\\
\hline
\hline
\end{tabular}\label{table_bestfits2}
}
\end{table}
\end{center}
\end{widetext}

The results after performing several running of the steps (a)-(b)-(c) above detailed\footnote{The codes are available in https://github.com/celia-escamilla-rivera/DEEoS-Cosmography} for the dark energy parameterisations in Sec. \ref{sec:models} are given in Tables \ref{table_bestfits} and \ref{table_bestfits2}. We notice that at this point we can easily perform the fit with model independent priors. Additionally, we compute the logarithm of the Bayes factor between two models $\mathcal{B}_{ij}=\mathcal{E}_{i}/\mathcal{E}_{j}$,
where the reference model ($\mathcal{E}_{i}$) with highest evidence is the $\Lambda$CDM model and with a flat prior over $H_0$. We apply the MCEvidence code\footnote{https://github.com/yabebalFantaye/MCEvidence} since it calculate the bayesian evolving from MCMC chains employed to fit the cosmological parameters $w_{i}$.
The interpretation scale known as Jeffreys's scale \cite{jeffreys}, is given as: if
$\ln{B_{ij}}<1$ there is not significant preference for the model with the highest evidence (or weak); if $1<\ln{B_{ij}}<2.5$ the
preference is substantial (or positive); if $2.5<\ln{B_{ij}}<5$ it is strong; if $\ln{B_{ij}}>5$ it is decisive (or very strong).

\begin{figure}
\centering
\includegraphics[width=0.48\textwidth,origin=c,angle=0]{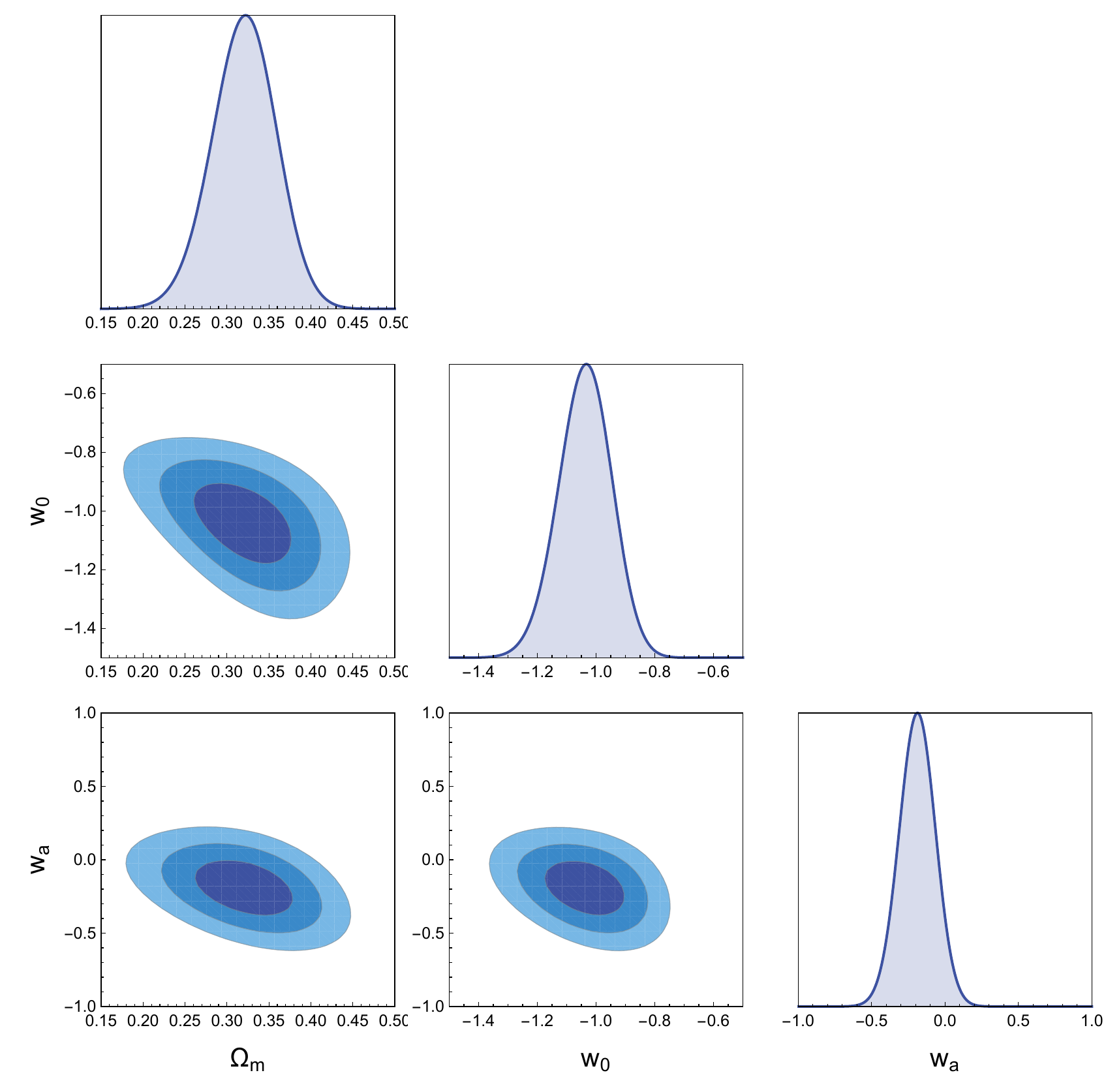}
\includegraphics[width=0.48\textwidth,origin=c,angle=0]{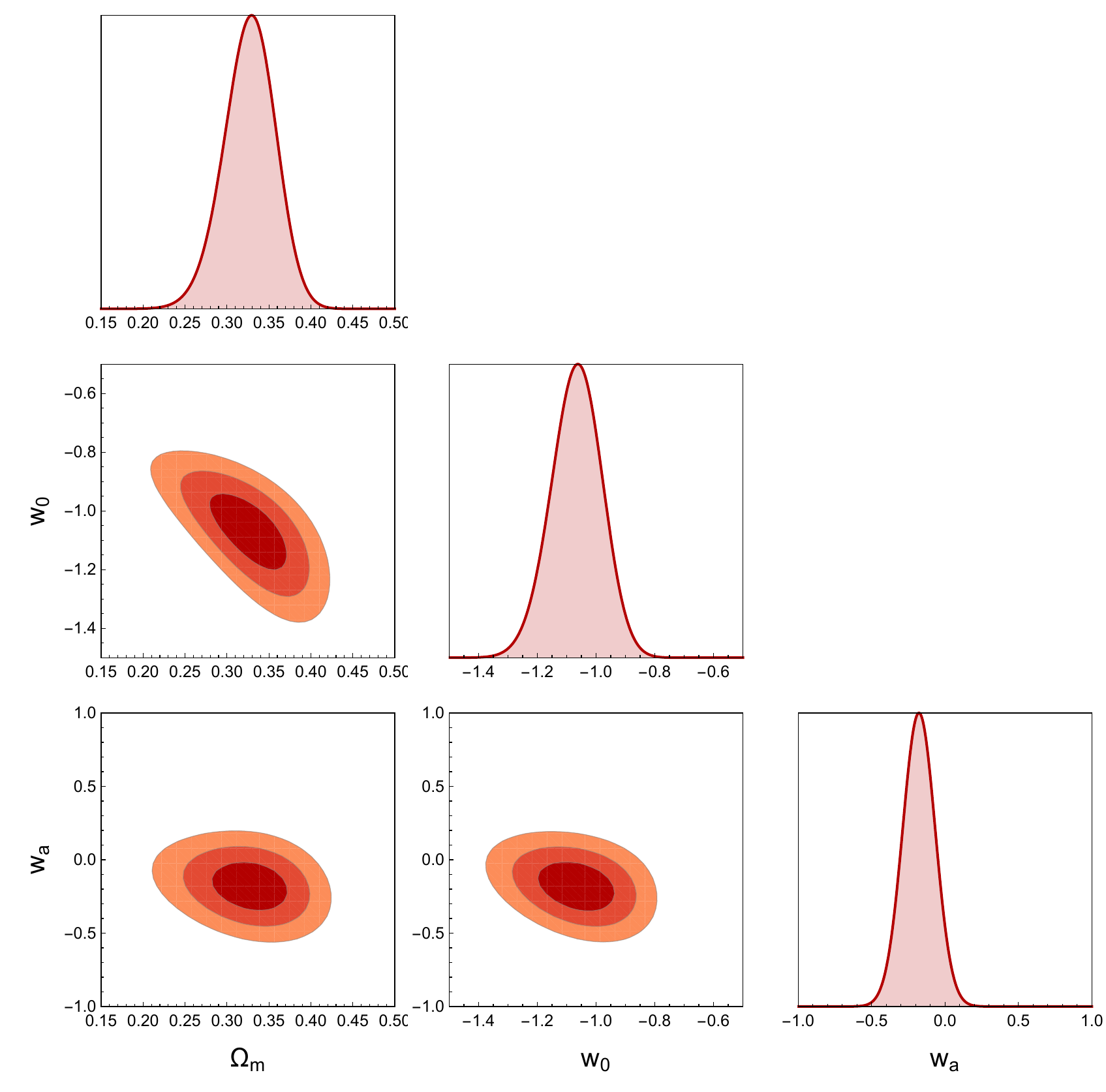} 
\caption{Marginalised posterior constraints for CPL parameterisation with 1$\sigma$ and 2$\sigma$ contours using JLA sample (top) and Pantheon sample (bottom) with model independent priors.} 
\label{fig:evolution_cpl_SNeIa}
\end{figure}

\begin{figure}
\centering
\includegraphics[width=0.48\textwidth,origin=c,angle=0]{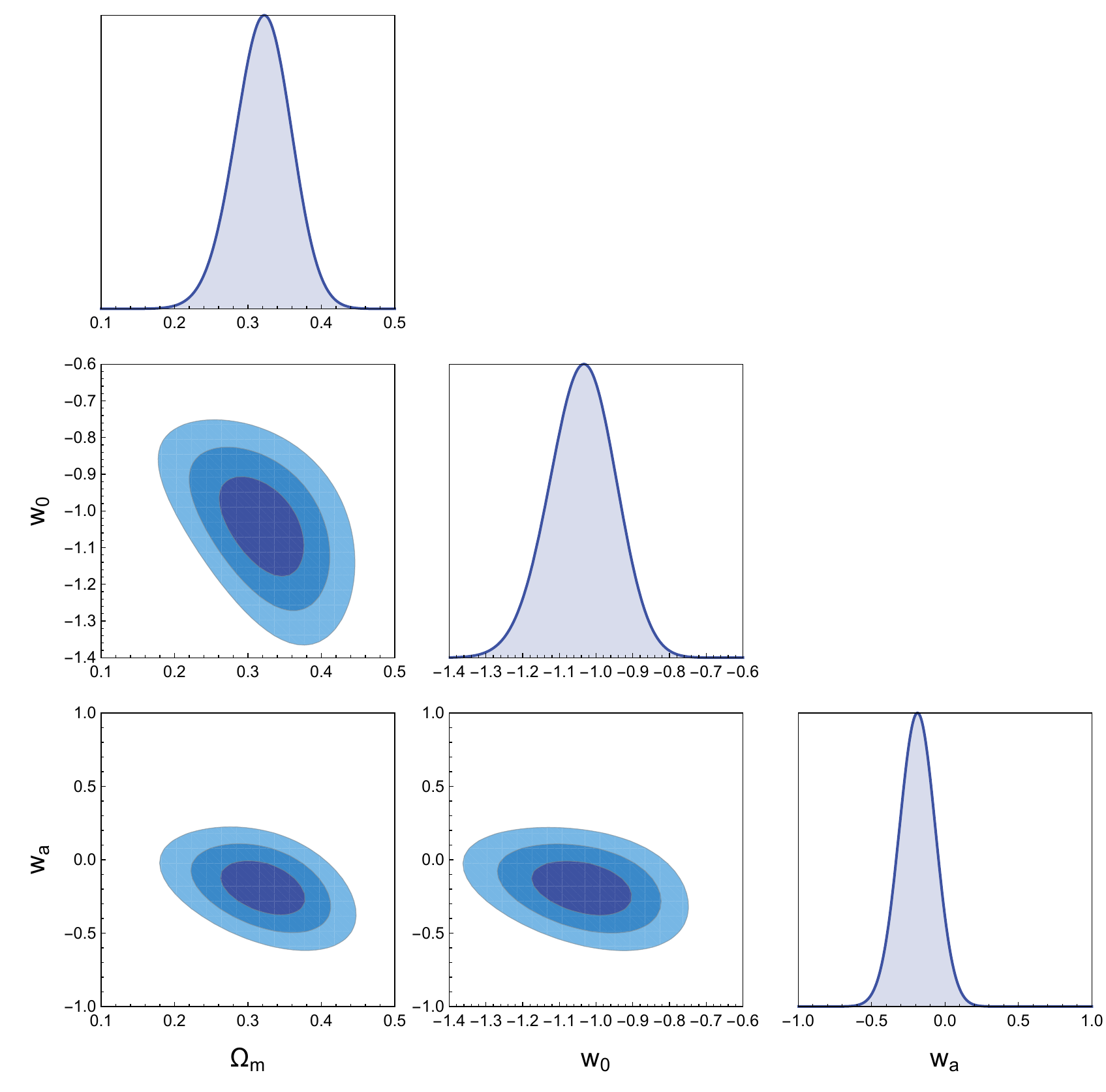}
\includegraphics[width=0.48\textwidth,origin=c,angle=0]{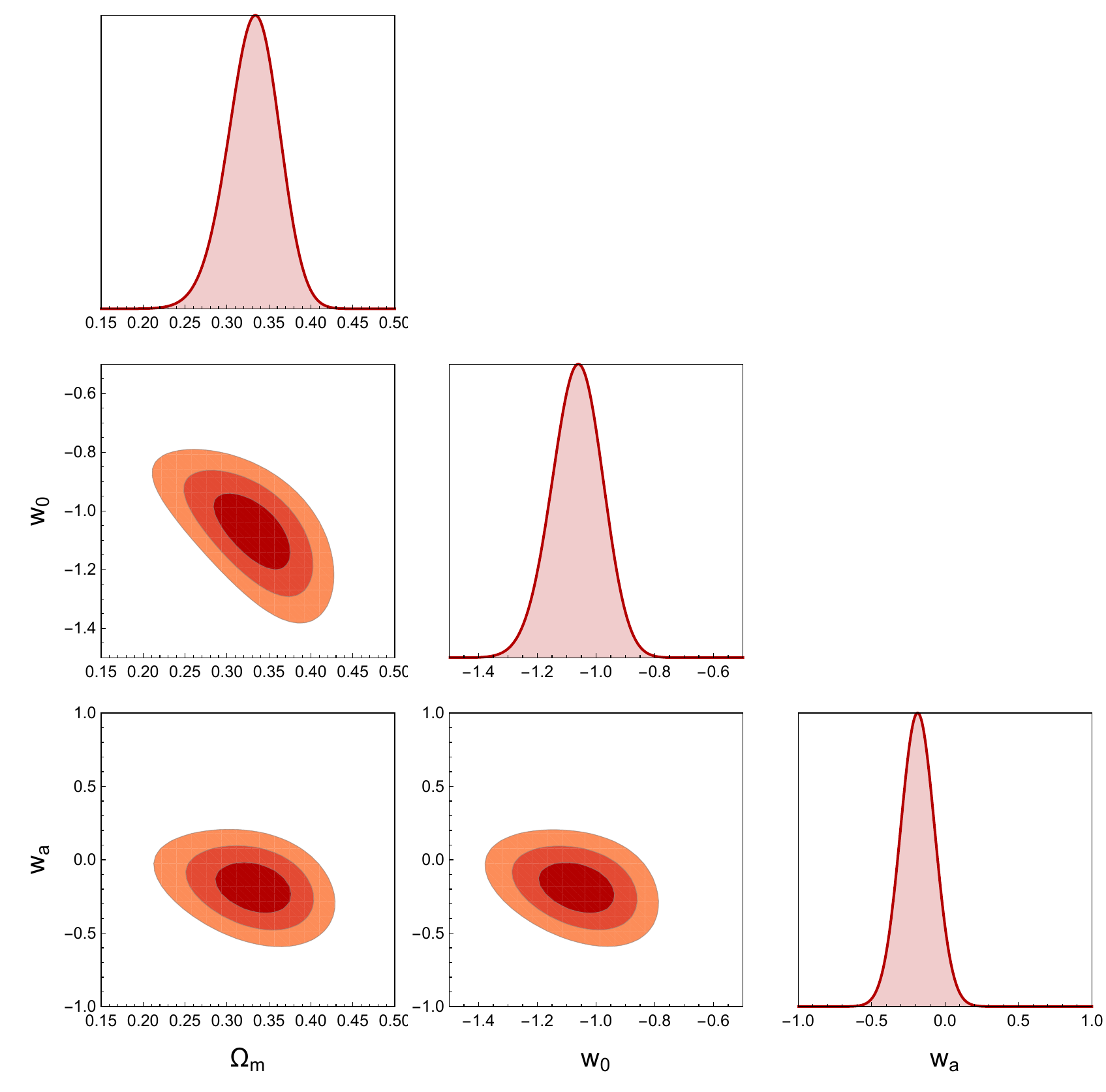} 
\caption{Marginalised posterior constraints for R-S parameterisation with 1$\sigma$ and 2$\sigma$ contours using JLA sample (top) and Pantheon sample (bottom) with model independent priors.} 
\label{fig:evolution_cpl_SNeIa}
\end{figure}


\section{Conclusions and remarks}
\label{sec:remarks}

The efficiency of the methodology presented strongly depends on the number of cosmographic parameters we are treating with and on how many EoS parameters we are going to consider free. For the sake of clarity, we have three possibilities:
\begin{enumerate}
\item EoS constant case and with two cosmographic parameters $[q_0,j_0]$: we can obtain information with $\Omega_m$ and $w_0=\text{const.}$
in terms of the cosmographic parameters, i.e $\Omega_m = \Omega_m (q_0,j_0)$ and $w_0=w_0(q_0,j_0)$.
\item Dynamical EoS case with two cosmographic parameters and not considering a prior over $\Omega_m$: we can obtain information in terms of  $w_0 =w_0 (q_0,\Omega_m)$ and $w_{i}= w_{i} (q_0,j_0,\Omega_m)$.
\item Dynamical EoS case with higher order cosmographic terms: we can obtain information over $\Omega_m=\Omega(q_0, j_0,s_0, l_0)$. And the same holds true for $w_{i} =w_{i}(q_0, j_0,s_0, l_0)$. 
\end{enumerate}
From this point forward, we can easily find out what are the cosmographic parameter values expected for the $w_x (z)$ in consideration by using the surveys of our preference. In the scenarios presented, we found that using supernovae samplers to test a specific cosmological model it is possible to \textit{compute directly the cosmographic parameters which are constrained by the bayesian evidence given by the model in question.} At this point, it is important to remark that in Busti et al 2016 and references therein \cite{Busti:2015xqa} was reported the impossibility of a cosmographic methodology than can provide reliable results to distinguish between models, but 
while, overall, the best fits parameters for each parameterisation shows the usual dark energy evolution, notice that \textit{with our presented proposal we can use
the statistical evidence\footnote{Although we used JLA and Pantheon samplers, we can easily add combined probes (Cosmic Clocks, BAO, GRB, etc. data sets) in our proposal. We leave these analyses for future work.} to set a cut off over the preference of one specific model and therefore with the posterior distribution distinguish between cosmological models.}

Interesting enough, in this paper we combine a couple of tools to select among competing standard dark energy models: our Eqs. (\ref{eq:newcosmoEoS})-(\ref{eq:q})-(\ref{eq:j}) and the bayesian selection criterion. On one hand, cosmography does not account for any model a priori, and it is so far the most powerful technique to derive cosmological bounds directly from astrophysical surveys. When we compare our results with recent supernovae Type Ia samplers, it is possible to show that the $\Lambda$CDM and $w$CDM models seem to be preferred in comparison to other dark energy models. But this does not imply that bidimensional parameterisations, as the ones presented in this work, can be competitive using future surveys. Even more, scenarios with cosmic fluids \cite{Capozziello:2013wha,Brevik:2017msy} can be consider in order to: \textit{(1)} found cosmological bounds by fitting the data and see if a particular model passes (or not) present cosmological constraints and \textit{(2)} discriminate this particular model against other models using the selection criterion.
Therefore, we found that the advantage to combine cosmography with bayesian selection criterion give us the possibility to relax disadvantageous degeneracy problems over the cosmographic parameters. Our proposal opens, in principle, a path that might help in addressing present and future alternative theories of gravity or even extended theories of gravity. This issue will be reported elsewhere in future works.

It is remarkable to point out that even very well fitted cosmological parameters can correspond to large uncertainties in the cosmographic ones, being larger for higher order terms. Finally, we want to stress again that the above approach is, in wide sense, model independent. The information that we can acquire on the underlying cosmology strictly depends on the accuracy of data without imposing arbitrary priors. In this sense, it can constitute  a pipeline to reconstruct the cosmic history at any redshift depending on  the  level of precision and the control of systematics in observations.  


\begin{acknowledgments}
CE-R is supported by the Royal Astronomical Society as FRAS 10147. SC is supported in part by the INFN sezione di Napoli, {\it iniziative specifiche} QGSKY and MOONLIGHT2. The article is also based upon work from COST action CA15117 (CANTATA), supported by COST (European Cooperation in Science and Technology).
\end{acknowledgments}


\end{document}